\documentclass[%
 reprint,
 amsmath,amssymb,
 aps,
pra,
]{revtex4-2}

\usepackage{graphicx}
\usepackage{tabularx}
\usepackage{dcolumn}
\usepackage{bm}
\usepackage[dvipsnames]{xcolor}

\begin{document}

\preprint{APS/123-QED}

\title{Membrane stepping  optimization in Modulation Based Imaging}


\author{Paola Perion\(^{1,2}\), Clara Magnin\(^{3,4}\), Fulvia Arfelli\(^{1,2}\), Bertrand Faure\(^{4}\), Ralf Hendrik Menk\(^{2,5,6}\), Emmanuel Brun\(^{3}\)}
\affiliation{\(^{1}\)University of Trieste, Trieste, Italy}
\affiliation{\(^{2}\)Istituto Nazionale di Fisica Nucleare, Division of Trieste, Trieste, Italy}
\affiliation{\(^{3}\)Univ Grenoble Alpes, Inserm Strobe, Grenoble, France
}%
\affiliation{\(^{4}\)Xenocs SAS, Grenoble, France}
\affiliation{\(^{5}\)Elettra Sincrotrone Trieste, Trieste, Italy}
\affiliation{\(^{6}\)Department of Computer and Electrical Engineering, Mid Sweden University, Sundsvall, Sweden}
\date{\today}

\begin{abstract}
Modulation-based imaging (MoBI) is an X-ray phase-contrast technique that uses an intensity modulator (or membrane) in the beam. Although MoBI can be performed in a single shot, multiple exposures are typically needed to improve the quality of the result. The membrane is typically moved using a regular stepping pattern for convenience; however, the impact of the membrane movement scheme on image quality has not been fully investigated yet. In this work, we explore optimized movement strategies aiming at improving MoBI performance. An experimental study tested optimization schemes based on global and local standard deviation metrics, and compared them with regular and random stepping motions. The results demonstrated superior contrast-to-noise ratio and reduced angular sensitivity in the optimized approaches compared to conventional stepping. These results were consistent across different membrane types, with honeycomb membranes showing the highest compatibility with the optimization procedure. Noise power spectrum analysis further validated the advantages of the optimized motion strategies. Overall, the results demonstrate that an optimized membrane movement can significantly improve MoBI image quality without increasing experimental complexity. 
\end{abstract}

\maketitle


\section{Introduction}

 X-ray Phase contrast imaging (XPCI) detects the phase shifts that occur in X-rays as they traverse matter \cite{bravin2012x,quenot2022x}. Unlike conventional absorption-based X-ray imaging, XPCI offers higher visibility of low-absorbing materials, such as soft tissues, as it is sensitive to the small variations in the refraction index.

Thanks to the availability of high-brilliance synchrotron sources and, more recently, advanced X-ray laboratory sources, it has become routine to observe and exploit refraction and interference phenomena in various fields \cite{walsh2021imaging,partridge2022enhanced}. As a consequence, X-ray phase effects in imaging are now effectively used in combination with traditional attenuation effects to provide complementary information about a sample's internal structure.

Various techniques exist to convert the phase shifts of X-rays into modulations of intensity recorded at the detector. Among them, \textit{Free-space propagation} is probably the simplest, requiring only an appropriate propagation distance between the sample and the detector to make phase effects detectable~\cite{snigirev1995possibilities, wilkins1996phase}. 

Another category of X-ray phase-contrast imaging techniques is based on the use of structured illumination to extract information about absorption, refraction and scattering. A known pattern, such as an absorption grid~\cite{vittoria2015beam} or an intensity modulator~\cite{morgan2012x,berujon2012x} is imposed on the beam. First, reference images of the known pattern are acquired with no sample. Then, after placing the sample, the altered wavefront is recorded. Phase shifts introduced by the sample distort the pattern, while absorption reduces its overall intensity. By tracking these modifications, it is possible to recover absorption, refraction, and dark-field images.

In general, modulation-based imaging (MoBI) is the term used when referring to XPCI performed using an intensity modulator (or membrane) placed in the beam. Among the existing membranes, sandpaper produces randomly distributed intensity patterns (or speckles if coherent light is employed), which generally have irregular shapes and random distribution of sizes. Small, well-defined speckles that cover a few pixels in the detector plane are usually preferred~\cite{zdora2018state}. Recently, alternatives to sandpaper membranes have been explored with the aim of optimizing membrane modulation patterns~\cite{magnin2504optimisation}. These include the use of structured membranes designed to produce specific hole patterns in different geometries.

The advantages of MoBI are the absence of intrinsic limitations in terms of field of view and spatial resolution beyond those imposed by the detector itself, as membranes of any size and granularity can be found. Experimental simplicity is achieved at the cost of increased numerical processing of the acquired data. Although MoBI can, in principle, be performed in a single exposure~\cite{pavlov2020single,ganz2022single}, image quality is typically improved by acquiring several pairs of reference and sample images while moving the membrane between exposures. In most implementations of MoBI, this shift is performed in a regular stepping pattern, mainly for practicality reasons. This shift can be performed either one- or two-dimensionally, depending on the experimental setup. Similar strategies are used in related techniques, such as beam tracking using 2D absorption grids, where the regular stepping naturally matches the periodic grid structure~\cite{navarrete2023two}. Alternative motions schemes have also been investigated: for example, spiral trajectory of sandpaper membranes have been employed to improve sampling across the field of view~\cite{savatovic2025high}. 

While some alternative motion schemes have been explored in the literature, the specific impact of different membrane movement patterns in MoBI remains unexplored. In this work, we investigate this to determine if, and to what extent, the choice of motion type affects the quality of the final image in MoBI reconstructions, with the aim of optimizing the movement strategy to maximize image quality.

\section{Materials and methods}
MoBI operates by creating a radiation field with a well-known structure. Reference images are first acquired without the sample, after which the sample is introduced into the beam and the altered wavefront is recorded. By comparing reference and sample images, absorption, phase, and dark field information can be retrieved.

In this work, the analysis was conducted experimentally to evaluate the feasibility of optimizing the membrane stepping procedure. The following sections describe the experimental setup, the stepping optimization procedure, and the final data analysis.


\subsection{Experimental Setup}
\label{Exp setup}
\begin{figure}[h]
\includegraphics[width=0.45\textwidth]{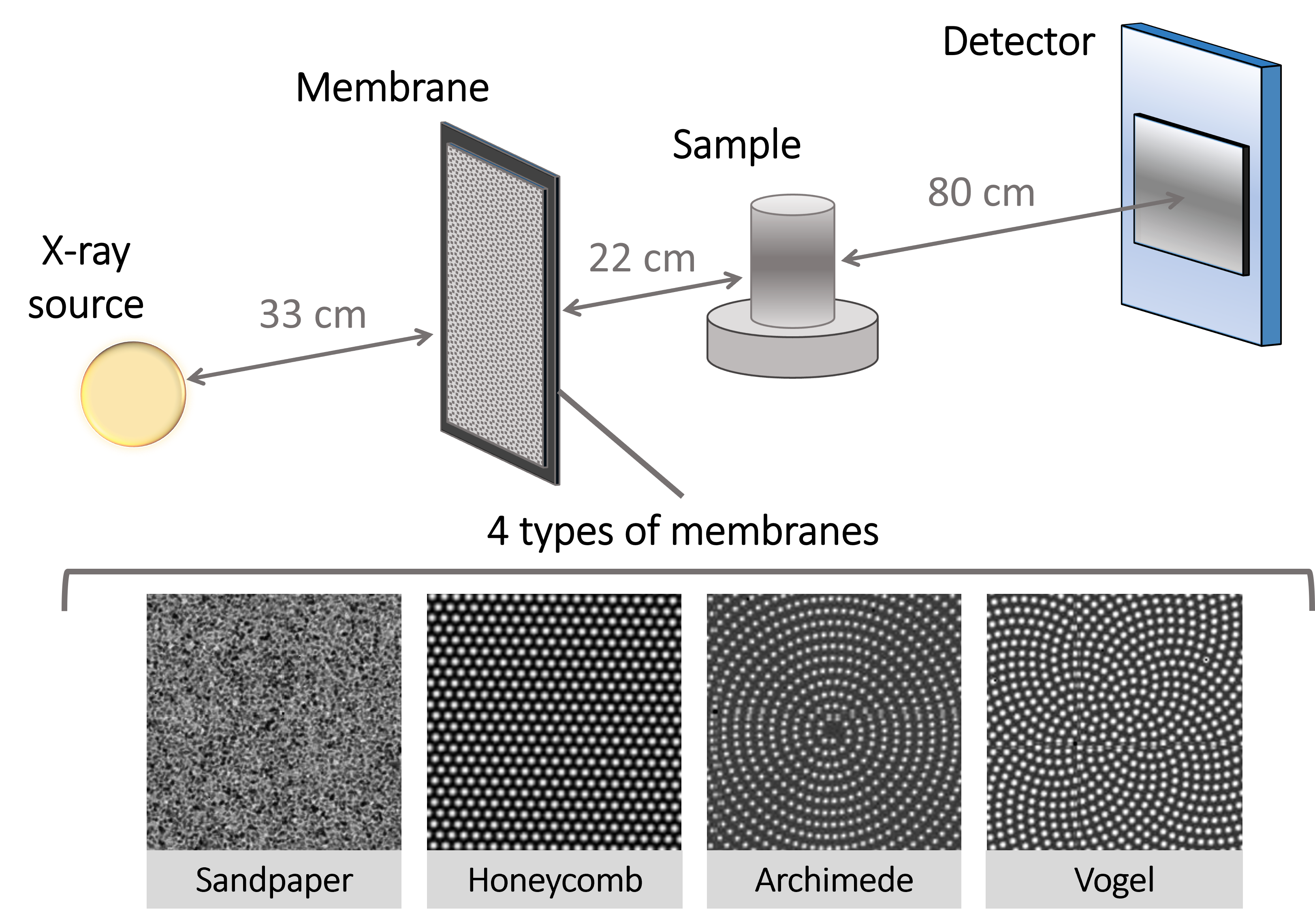}
\caption{\label{Figure1} Experimental setup sketch showing the relevant distances.}
\end{figure}

The experimental setup is shown in Figure~\ref{Figure1}. Measurements were performed at Xenocs, (Xenocs SAS Grenoble, France), using the Xeuss 3.0 instrument \footnote{ https://www.xenocs.com/saxs-products/saxs-equipment-xeuss/}. For this work, its polychromatic cone-beam imaging  source, featuring a copper target and operating at 30kVp with 8.6 keV mean energy, was used.\\
Four different types of membrane were tested in the setup: 
\begin{enumerate}
    \item Sandpaper membrane
    \item Archimede: nickel membrane featuring holes arranged in an Archimedean spiral pattern
    \item Honeycomb: nickel membrane with holes organized in a hexagonal matrix
    \item Vogel: nickel membrane with holes arranged in a spiral pattern that mimics the arrangement of sunflower seeds \cite{vogel1979better}.
\end{enumerate}
The membrane patterns can be seen in Figure~\ref{Figure1}. The membrane were obtained by laser micro drilling using a fibre laser and was processed by STPgroup (STPgroup, Saint Ismier, France). 

The membranes were made of Ni foil of 30 $\mu m$ in thickness. These membranes were then drilled with a circular apertures of 80 $\mu m$ in diameter with a laser.

The scanned sample is a 3D-printed object that follows Julia's fractal geometry, fabricated in resin (standar photopolymer resin 8k from Elegoo, Shenzhen China) with dimensions 18x10 mm and maximum height of 5 mm. The in-plane resolution used was 18 $\mu m$ below the spatial resolution of our system. This sample was chosen because it presents features of many different sizes, allowing a better evaluation of the quality of the image.

The detector used was the Eiger2 500k from Dectris, which has a 1030x514 matrix of 75$\mu m$ pixels. The membranes, sample, and detector were placed in the primary vacuum. The relevant distances in the setup can be seen in Figure~\ref{Figure1}.

A total of 200 reference images ($I_r$) and 200 sample images ($I_s$) were acquired for each membrane while the membrane was stepped in the vertical direction with respect to the X-ray propagation axis.

\subsection{Stepping optimization procedure}

A dataset of 200 pairs of reference and sample images was acquired for each membrane, covering a wide range of stepping positions with step size of 70$\mu m$. This large dataset allowed for the selection of subsets that simulate different membrane movement patterns, enabling a comparative analysis of how different stepping procedures affect image quality.

To do this, four distinct stepping strategies were tested (see Figure~\ref{Figure2}). For each of these, subsets of N images were extracted from the full set of 200 reference images:

\begin{figure*}
\includegraphics[width=0.8\textwidth]{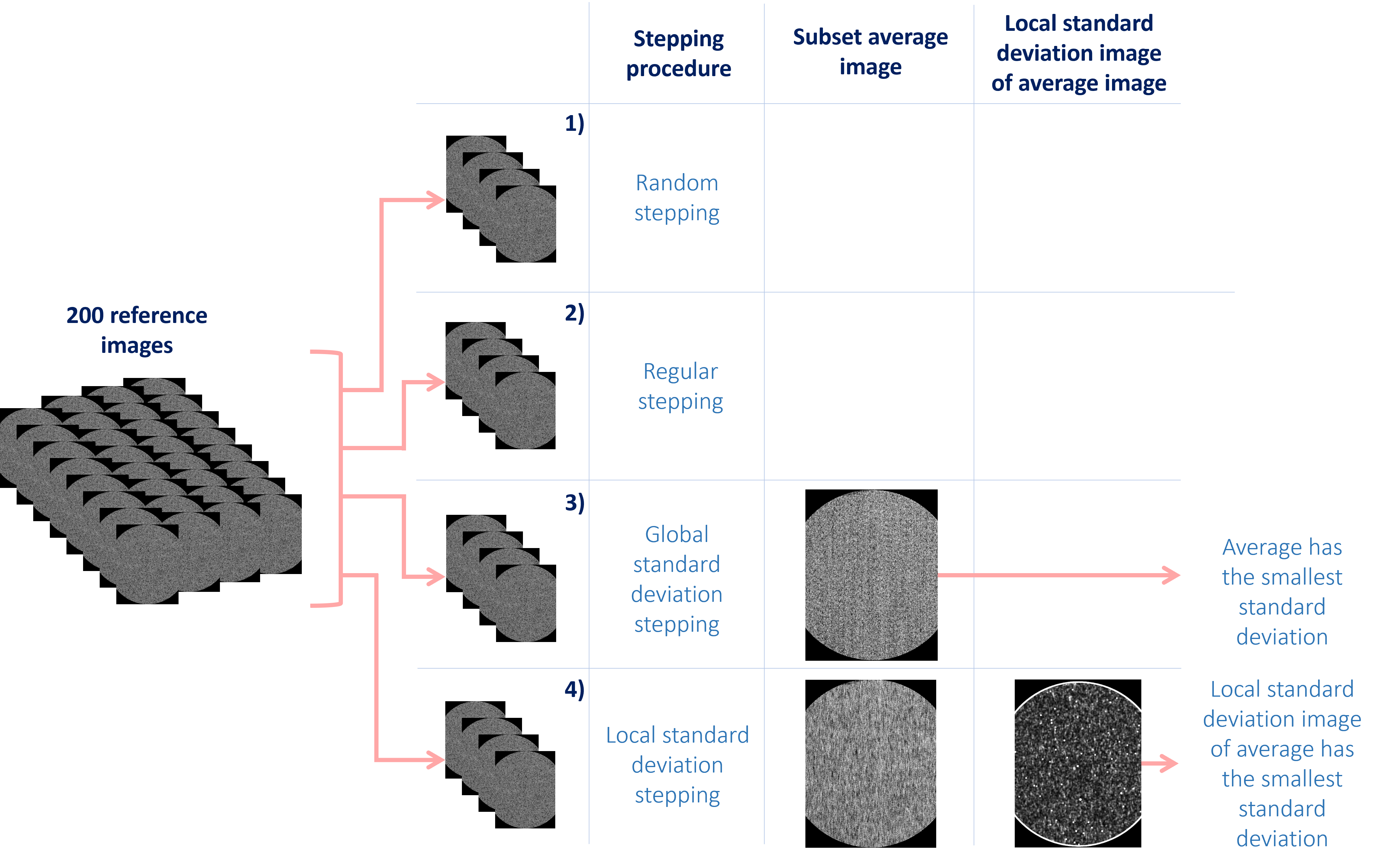}
\caption{\label{Figure2} Stepping optimization procedure. For each membrane type, 200 reference/sample image pairs were acquired. Subsets of size N were extracted using different stepping methods: subsets (1) are extracted randomly, subsets (2) are composed of consecutive images, subsets (3) were chosen by minimizing the standard deviation of their average image, and subsets (4) were selected by averaging the images, thus minimizing the standard deviation of their local standard deviation.}
\end{figure*}

\begin{enumerate}
  \item Random stepping: N membrane images were randomly chosen from the large dataset to simulate a random movement of the membrane.
  \item Regular stepping: N consecutive membrane positions were selected, mimicking a regular, stepwise membrane motion.
  \item Global standard deviation minimization: the subset of N images whose sum had the lowest global standard deviation was selected, with the goal of achieving the lowest global standard deviation across the entire image.
  \item Local standard deviation minimization: This method focuses on local uniformity. The sum of N selected membranes was computed, and a local standard deviation map was generated over 7×7 pixel regions. The subset of N membranes that exhibited the smallest standard deviation of this map was chosen. The aim was to avoid localized intensity fluctuations.
\end{enumerate}

In both optimization-based methods (3 and 4), the objective was to maximize illumination uniformity on the sample, both globally and locally, using standard deviation as an optimization metric to quantify intensity variation. A lower standard deviation in the sum image indicates an overall more homogeneous illumination. The local standard deviation stepping method was thought to prevent small pixel regions with high intensity variation.


\subsection{Data analysis}

The stepping optimization procedure described in the previous section was applied for each of the four membrane types shown in Figure~\ref{Figure1}. For every combination of membrane type and stepping strategy, seven subsets were extracted from the full set of 200 reference/sample pairs of images. Each subset contained a number N of pairs of reference/sample images, with N = 4, 5, 7, 10, 15, 20 and 30. This range was chosen to evaluate how the number of stepping positions affects image quality across different membrane types and stepping methods. This data handling is summarized in Table \ref{tab:stepping_overview}.

\begin{table*}[ht]
\centering
\caption{Overview of the data handling. For each of the four membrane types, a dataset of 200 reference/sample image pairs was acquired. From each dataset, subsets of size \(N = 4, 5, 7, 10, 15, 20, 30\) were extracted using four different stepping strategies.}
\label{tab:stepping_overview}
\begin{tabularx}{\textwidth}{>{\centering\arraybackslash}X>{\centering\arraybackslash}X>{\centering\arraybackslash}X>{\centering\arraybackslash}X}
\textbf{Membrane Type} & \textbf{Total Images} & \textbf{Stepping Strategies} & \textbf{Subset Sizes (\(N\))} \\
\hline
Archimede & 200 pairs & Random, Regular, Global Std, Local Std & 4, 5, 7, 10, 15, 20, 30 \\
Honeycomb & 200 pairs & Random, Regular, Global Std, Local Std & 4, 5, 7, 10, 15, 20, 30 \\
Sandpaper & 200 pairs & Random, Regular, Global Std, Local Std & 4, 5, 7, 10, 15, 20, 30 \\
Vogel & 200 pairs & Random, Regular, Global Std, Local Std & 4, 5, 7, 10, 15, 20, 30 \\
\end{tabularx}
\end{table*}

For each subset, absorption, refraction, and dark field images were recovered from reference and sample images using the implicit Low Coherence System (LCS) algorithm~\cite{magnin2023dark,quenot2021evaluation} based on Fokker-Planck equations~\cite{ morgan2019applying, paganin2019x}. In general, modulation-based imaging algorithms compare reference and sample images by analyzing how the reference pattern is distorted when the sample is inserted. LCS is based on flux conservation: the model relates the reference and sample images to the absorption $I_{obj}$, the displacement fields $D_x$ and $D_y$ (representing refraction), and a diffusion term $D_f$ (corresponding to the dark field signal) through the following equation:

\begin{eqnarray}
    I_r(x,y) = \frac{1}{I_{obj}(x,y)}I_s(x,y)+D_x(x,y)\frac{\partial I_r(x,y)}{\partial x}+\nonumber\\
    D_y(x,y)\frac{\partial I_r(x,y)}{\partial y}-z_2^{2}D_f(x,y)\nabla_\perp^2[I_r(x,y)]
\end{eqnarray}

where $I_r$ is the reference image, $I_s$ is the sample image and $z_2$ is the sample-to-detector distance. The equation presents four unknowns, indicating that a minimum of four membrane positions is required in order to solve the system of equations.

Once absorption, refraction, and dark-field images are retrieved, image quality is evaluated using the contrast-to-noise ratio (CNR) for absorption and dark field images and angular sensitivity for refraction images. The total CNR is calculated as the average CNR measured across six different 30x30 pixel regions inside the sample and six 30x30 pixel regions in the background:

\begin{equation}
    CNR = \frac{\bar\mu_{s} - \bar\mu_{bg}}{\bar\sigma_{bg}}
\end{equation}

where $\bar\mu_{s}$ and $\bar\mu_{bg}$ are the means of the individual mean intensity values of the regions respectively inside the sample and in the background, and $\bar\sigma_{bg}$ is the mean standard deviation of the regions of pixels in the background. The associated error is evaluated through propagation of the error. Multiple regions were chosen to give a more global estimate of image quality.

The angular sensitivity $\sigma_\theta$ is assessed in a region of the image without the sample by calculating the mean and standard error of the standard deviation of the refraction angles in twelve 45×45 pixel regions ($\sigma_{\theta_i}$):

\begin{equation}
    \sigma_{\theta} = \frac{1}{12}\sum_{i=1}^{12}{\sigma_{\theta_i}}
\end{equation}

Quantifying the dispersion of noise, the angular sensitivity determines the minimum detectable refraction angle of the system.

Finally, spatial resolution was calculated with an approach similar to that described in~\cite{modregger2007spatial}. The method determines spatial resolution by identifying the point in the frequency domain at which the signal becomes indistinguishable from noise. The noise power spectrum is calculated from the 2D image as a function of spatial frequency, the noise baseline is then identified from high-frequency components where only noise exists, and the resolution is calculated from the maximum spatial frequency where the total spectral power equals twice the noise baseline.

These quality metrics were evaluated for each membrane type, stepping method, and subset with a number of N reference/sample pairs. \\

\section{Results and Discussion}

\begin{figure}
\includegraphics[width=0.49\textwidth]{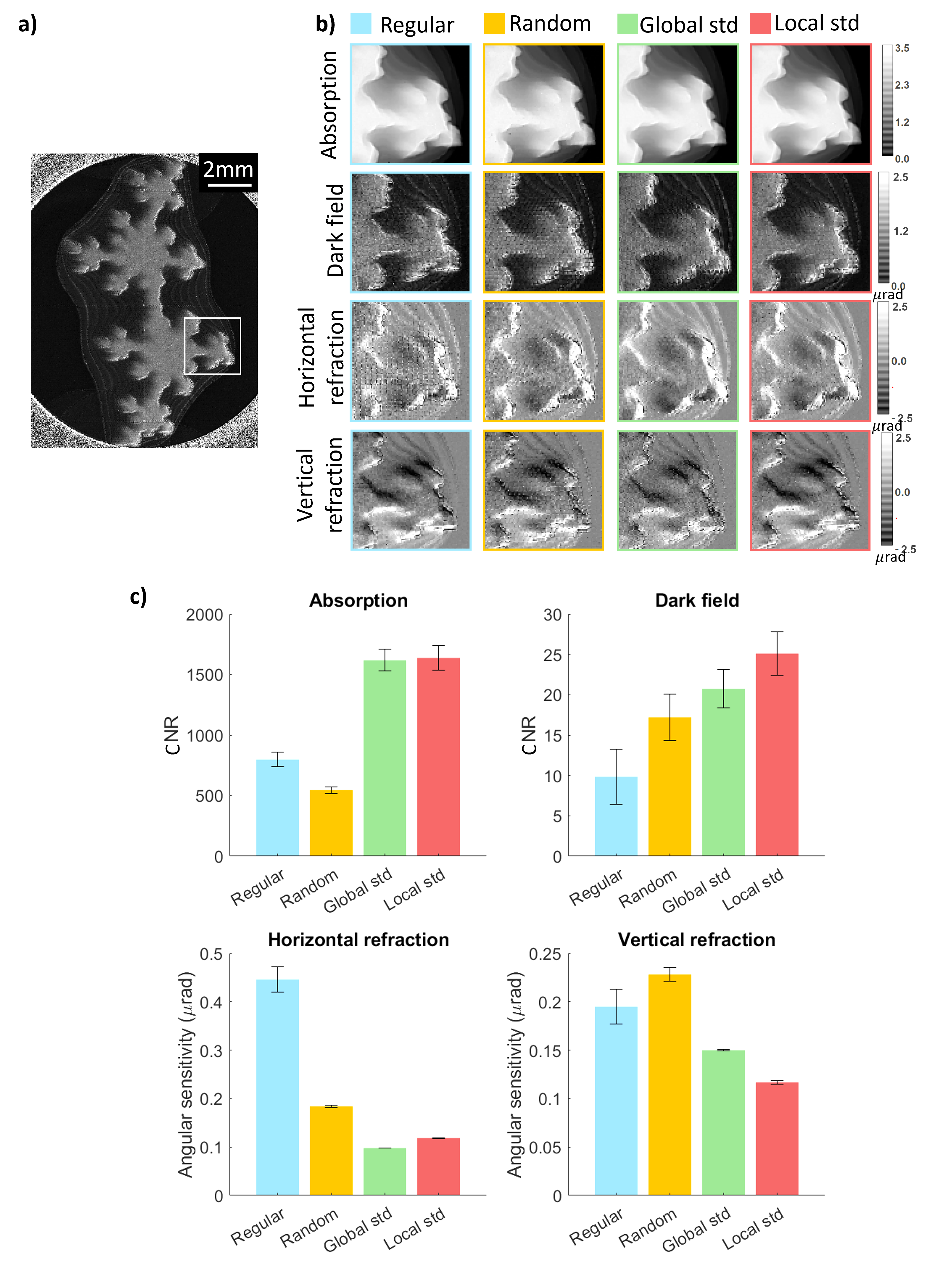}
\caption{\label{Figure4} a) Dark field image of the full sample obtained with a subset size of 5 and the honeycomb membrane. b) Absorption, dark field, horizontal refraction, and vertical refraction images retrieved using each optimization procedure: regular, random, global standard deviation, and local standard deviation. Only a cropped region of the image is shown for clarity. c) Quantitative evaluation: graphs displaying the quality metrics measured from the images as a function of the membrane stepping method.}
\end{figure}

The experiment involved four different membranes types \cite{magnin2504optimisation}, aiming not only to validate the proposed optimization approach but also to assess its consistency across various membrane types. Subsets containing 4, 5, 7, 10, 15, 20, and 30 membrane positions each were extracted from each dataset of 200 reference/sample pairs, using each of the four stepping procedures (regular, random, global standard deviation and local standard deviation). This allowed for an evaluation of how the number of stepping positions affects image quality.\\
Figure~\ref{Figure4} shows the results obtained using the honeycomb membrane and a subset size of 5 positions. The graphs display the quality metrics measured from the images as a function of the membrane stepping method. A cropped region of the image for each stepping method is displayed for visual comparison.

The results indicate that the global standard deviation and the local standard deviation stepping procedures generally produced the highest image quality whatever the modulation topology. Specifically, these methods achieved the highest CNR values both in absorption and dark field images, as well as the lowest angular sensitivity values in refraction. For both vertical and horizontal refraction, angular sensitivity values around 0.1$\mu rad$ were achieved with the optimized stepping methods, approximately twice as low as those obtained with regular and random stepping.

\begin{figure}
\includegraphics[width=0.49\textwidth]{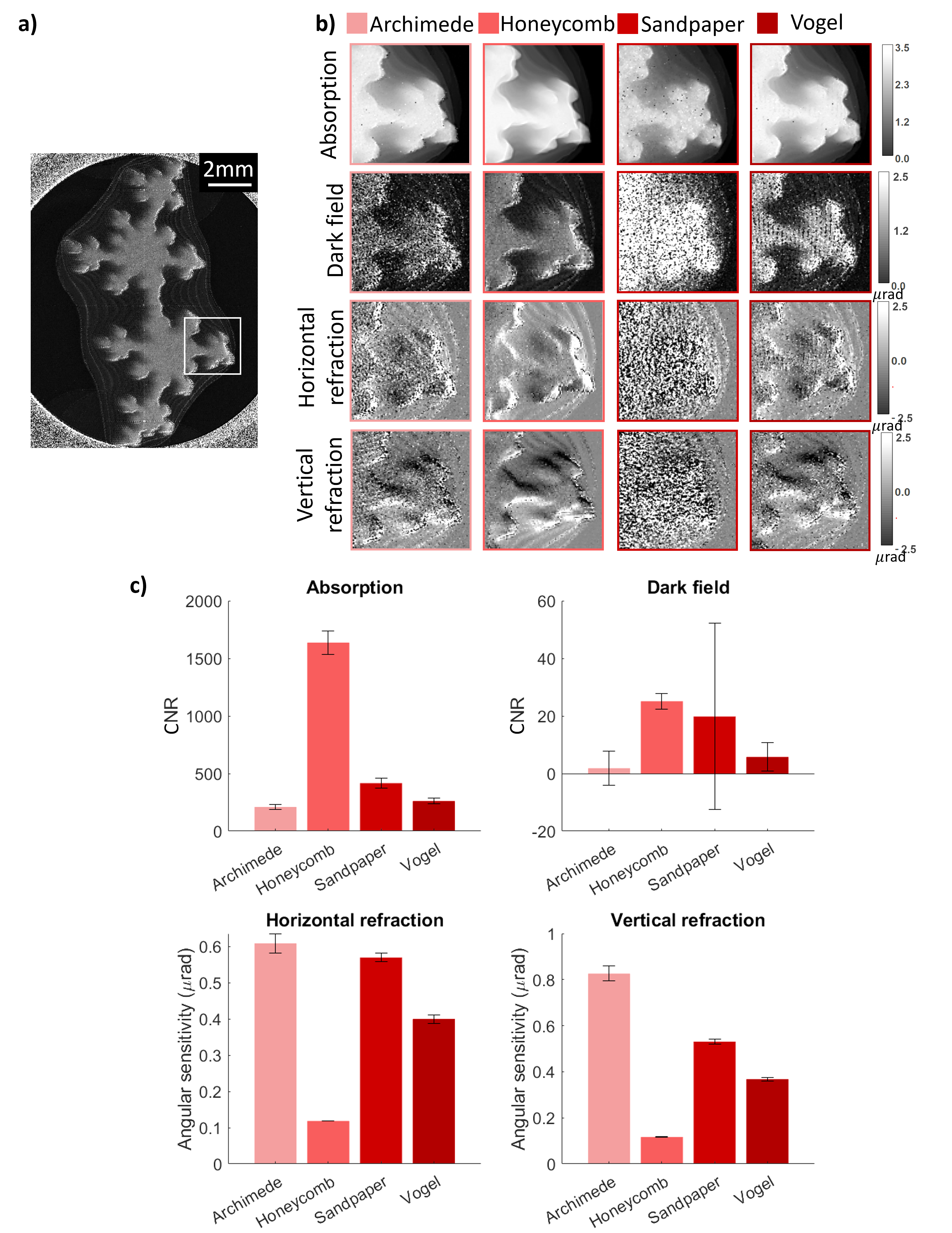}
\caption{\label{Figure5} a) Dark field image of the full sample obtained with a subset size of 5 and the local standard deviation method. b) Absorption, dark field, horizontal refraction, and vertical refraction images retrieved using each membrane: archimede, honeycomb, sandpaper, and Vogel. Only a cropped region of the image is shown for clarity. c) Quantitative evaluation: graphs displaying the quality metrics measured from the images.}
\end{figure}

Given its overall better performance, the local standard deviation method was selected to evaluate the performance of the different membrane types. The results obtained using this method for each membrane, with a subset size of 5, are presented in Figure~\ref{Figure5}. The same quality metrics were computed and plotted as a function of the membrane type. Again, only a cropped region of the image is displayed for visual comparison.\\

The results indicate that the honeycomb membrane consistently yields the optimal results, with the highest CNR values in both absorption and dark field images, as well as the lowest angular sensitivity in refraction images. Specifically, the honeycomb membrane achieves angular sensitivity values around 0.1$\mu rad$, whereas the other membranes range between 0.3$\mu rad$ and 0.8$\mu rad$. This suggests stronger compatibility between the honeycomb membrane and the optimization procedure. It is also worth noting that the sandpaper membrane produced a high CNR value in the dark-field image, primarily because of the strong signal output image. Moreover, the noise estimation in the background was reliable for all membranes except sandpaper, where the noise within the sample was significantly higher. \\

\begin{figure*}
\includegraphics[width=0.8\textwidth]{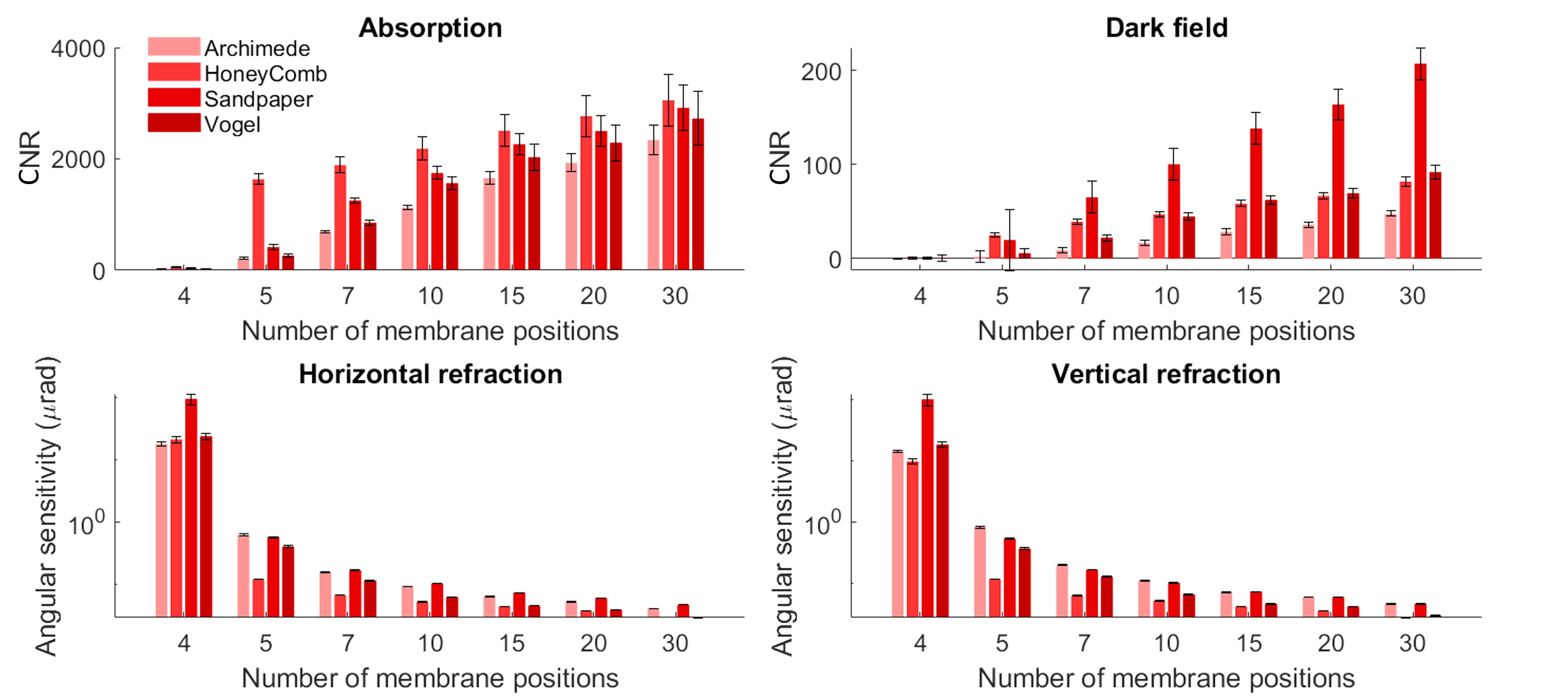}
\caption{\label{Figure6} Quality metrics measured from the absorption, refraction and dark-field images for each membrane type, plotted as a function of the number of membrane positions included in the subset.}
\end{figure*}

The results presented above are all obtained from a fixed subset size of 5 pairs of sample/membrane images. However, as mentioned earlier, the study was conducted across multiple subsets sizes. Figure~\ref{Figure6} presents the results obtained for each membrane using the local standard deviation procedure, showing how the quality metrics vary with the number of membrane positions included in each subset.

\begin{figure}
\includegraphics[width=0.45\textwidth]{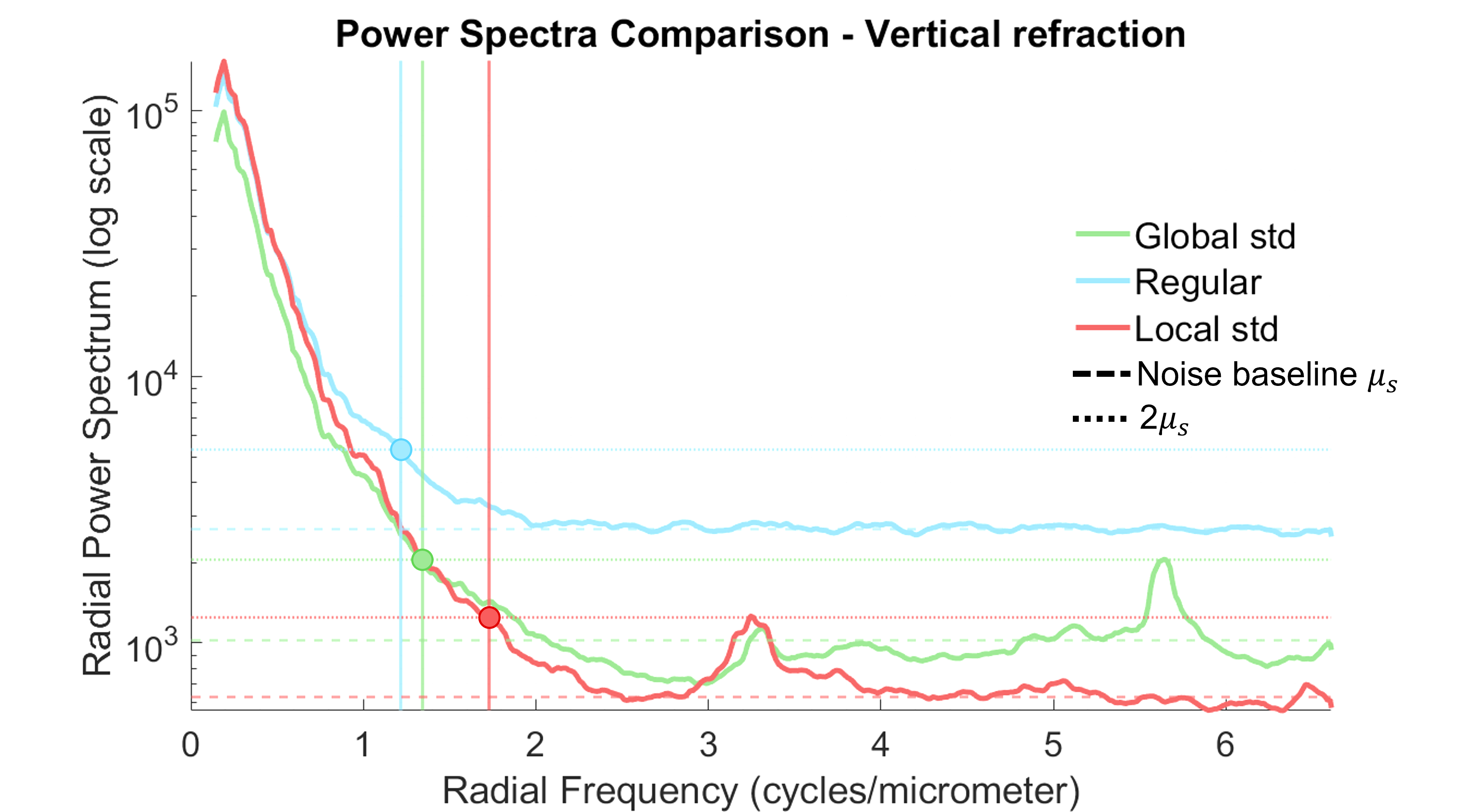}
\caption{\label{Figure7} Radial noise power spectrum calculated from vertical refraction images for each membrane stepping method. The random stepping method is not shown as it did not converge. The noise baseline is identified from high-frequency components where only noise exists, and the resolution is calculated from the maximum spatial frequency where the total spectral power equals twice the noise baseline (dots on spectrum).}
\end{figure}

Again, the honeycomb membrane delivers the best overall performance, showing a rapid increase in CNR when the number of membrane positions increases from 4 to 5 in the absorption images. The same trend is seen in angular sensitivity as the number of positions increases. These trends are observed across all membranes types, although with a more gradual improvement in quality metrics. Notably, for the dark field CNR, the honeycomb membrane exhibits a rapid initial increase, but the Vogel membrane outperforms it when more than 10 membrane positions are used. It is worth noting that directional darkfield \cite{magnin2504optimisation} was not tested in this article.\\
Finally, spatial resolution was evaluated through the noise power spectrum method. Figure~\ref{Figure7} shows the noise power spectra derived from the vertical refraction images, using a subset of 5 honeycomb membrane positions. In the plot, the noise power spectrum obtained for the random stepping is not showed as it did not converge. The corresponding resolution values, reported in Table~\ref{tab:spatialres}, are reported for horizontal and vertical refraction, and dark-field, allowing a comparison of the performance across different membrane stepping methods.

\begin{table}[h]
\centering
\caption{Spatial resolution values for each membrane stepping method across refraction and dark-field images. The value for the random stepping method in the case of vertical refraction is not displayed as it did not converge. HR: Horizontal Refraction; VR: Vertical Refraction; DF: Dark Field}
\label{tab:spatialres}
\small
\begin{tabularx}{0.48\textwidth}{l*{3}{>{\centering\arraybackslash}X}}
\textbf{Method} & \textbf{HR} & \textbf{VR} & \textbf{DF} \\
& \textbf{(mm)} & \textbf{(mm)} & \textbf{(mm)} \\
\hline
Random & 4.17 $\pm$ 0.56 & - & 3.68 $\pm$ 0.43\\
Regular & 3.29 $\pm$ 0.35 & 0.823 $\pm$ 0.033 &  1.16 $\pm$ 0.09\\
Global std & 0.504 $\pm$ 0.069 & 0.745 $\pm$ 0.198 & 0.544 $\pm$ 0.264\\
Local std & 0.665 $\pm$ 0.056 & 0.579 $\pm$ 0.049 & 0.554 $\pm$ 0.020\\
\end{tabularx}
\end{table}

The noise power spectra clearly show the superiority of the global and local standard deviation stepping methods compared to the random and regular approaches. Both the local and global methods achieve sub-millimeter spatial resolution across all modalities, supporting the effectiveness of the proposed stepping optimization strategy. The global standard deviation method exhibits greater variability, indicating a higher sensitivity to local fluctuations. In contrast, the regular and random methods yield poorer resolutions, ranging from approximately 1 mm to over 4 mm. Notably, the random method fails to converge for vertical refraction.

The findings of this work suggest the advantage of optimizing the membrane stepping strategy at the beginning of an experiment. By initially sampling a large number of different membrane positions, the local or global standard deviation methods can be applied to identify the optimal stepping pattern. Once this is determined, the corresponding motor positions can be recorded and used consistently throughout the experiment. This procedure ensures that data acquisition is performed under optimized and reproducible conditions, ensuring an optimized and stable image quality.

\section{Conclusions}

This work demonstrated the effectiveness of optimized movement strategies in MoBI through an experimental study. The results show that the two optimized stepping procedures consistently produced the highest image quality, achieving the highest CNR in the absorption and dark field images, the best angular sensitivity in refraction images, and superior spatial resolution overall.

The study was performed using four different types of membranes: the honeycomb membrane yielded the best results overall, suggesting a stronger compatibility with the optimization procedure compared to the other membranes. In addition, the quality analysis conducted using the noise power spectrum method confirmed the superiority of both the global and local standard deviation stepping methods.

Future work may extend this study exploring alternative optimization procedures, also including a broader range of membrane types or different experimental conditions including directional dark-field sensitivity. Ultimately, the optimization of membrane movement in MoBI has the potential of enhancing image quality without increasing experimental complexity, thereby strengthening the technique's overall performance and potentially reducing the number of exposures required to achieve the same image quality.

\begin{acknowledgments}
Part of this work was supported by the LABEX PRIMES (ANR-11-LABX-0063) of Université de Lyon, within the program  "Investissements d’Avenir"  (ANR-11-IDEX-0007) operated by the French National Research Agency (ANR). We acknowledge the support of ANRT cifre n°2022/0476 program that partly funds C.M. E.B acknowledges the support of Inserm IRP linx project.\\
Part of this work was supported by the Erasmus+ Traineeship program from the University of Trieste. F.A. acknowledges the support of INFN group V.
\end{acknowledgments}

\nocite{*}
\bibliography{apssamp}

\end{document}